\documentclass[aps,prb,reprint,superscriptaddress,longbibliography]{revtex4-1}%

\usepackage{graphicx}
\usepackage{amssymb, amsmath}
\usepackage[usenames]{color}
\usepackage{dcolumn}
\usepackage{bm}
\usepackage{ textcomp }
\usepackage[normalem]{ulem}

\begin{document}
\title{Phononic Stiefel-Whitney topology with corner vibrational modes in two-dimensional Xenes and ligand-functionalized derivatives}

\author{Mingxiang Pan}
\affiliation{School of Physics, Peking University, Beijing 100871, China}

\author{Huaqing Huang}
\email[Corresponding author: ]{huaqing.huang@pku.edu.cn}
\affiliation{School of Physics, Peking University, Beijing 100871, China}
\affiliation{Collaborative Innovation Center of Quantum Matter, Beijing 100871, China}
\affiliation{Center for High Energy Physics, Peking University, Beijing 100871, China}

\date{\today}

\begin{abstract}
Two-dimensional (2D) Stiefel-Whitney (SW) insulator is a fragile topological state characterized by the second SW class in the presence of space-time inversion symmetry. So far, SWIs have been proposed in several electronic materials but seldom in phononic systems. Here we recognize that a large class of 2D buckled honeycomb crystals termed Xenes and their ligand-functionalized derivatives realize the nontrivial phononic SW topology. The phononic SWIs are identified by a nonzero second SW number $w_2=1$, associated with gaped edge states and robust topological corner modes. Despite the versatility of electronic topological properties in these materials, the nontrivial phononic SW topology is mainly attributed to the double band inversion between in-plane acoustic and out-of-plane optical bands with opposite parities due to the structural buckling of the honeycomb lattice. Our findings not only reveal an overlooked phononic topological property of 2D Xene-related materials, but also afford abundant readily synthesizable material candidates with simple phononic spectra for further experimental studies of phononic SW topology physics.
\end{abstract}


\maketitle

\section{Introduction}
The rise of higher-order topological insulators (HOTIs) over the past few years has inspired the interests of the modified bulk-boundary correspondence where the nontrivial bulk topological invariant in $d$ dimension guarantees the existence of topological boundary states not at $(d-1)$- but at $(d-n)$-dimensional boundaries with $n > 1$.\cite{benalcazar2017quantized,schindler2018higher,PhysRevLett.120.026801, PhysRevLett.119.246401, PhysRevLett.119.246402} One closely related class of HOTIs are the Stiefel-Whitney insulators (SWIs), which are characterized by the second SW number $w_2$ in spin-orbit-free systems respecting the combination of spatial and time reversal symmetry ($\mathcal{PT}$).
\cite{PhysRevLett.116.156402,PhysRevLett.118.056401,PhysRevLett.125.126403, PhysRevLett.121.106403, Ahn_2019, PhysRevX.9.021013}
So far, 2D SWIs have been proposed in electronic materials where spin-orbit coupling (SOC) is negligibly weak, such as twisted bilayer graphene, \cite{PhysRevLett.123.216803,PhysRevLett.126.066401} graph(di)yne, \cite{liu2019two,lee2020two,PhysRevLett.123.256402, PhysRevB.104.085205, PhysRevB.106.035153} 2D group-V materials,\cite{PhysRevB.98.045125,PhysRevB.102.115104,PhysRevB.104.245427,PhysRevB.105.075407,PhysRevResearch.1.033074,huang2021structrual} and ligand-functionalized Xenes.\cite{huanghq2022pan, PhysRevB.104.245427}
Strictly speaking, SOC is inevitable in electronic materials, therefore, the spin-orbit-free requirement, which is a stringent condition for electronic SWI, can be inherently avoided in phononic systems due to the spinless feature of phonons. However, material candidates for phononic SWIs are still very rare up to now,\cite{acs.nanolett.1c04239,PhysRevB.105.085123} which poses a great challenge for the experimental study of topological physics associated with SW class.

Encouraged by the great success of graphene, a large class of 2D monoelemental materials (named Xenes) \cite{molle2017buckled,zhao2020two,BECHSTEDT2021100615,C7CS00125H,acsnano.7b07436} have gained tremendous interest due to their intriguing electronic properties including various types of topological states induced by chemical functionalization, \cite{PhysRevLett.111.136804,PhysRevLett.107.076802,PhysRevLett.113.256401,PhysRevLett.109.055502,PhysRevB.98.045125,huanghqCMS} semiconductors with tunable gaps,\cite{PhysRevLett.111.136804,PhysRevB.89.115429,nl5025535} and Dirac semimetals.\cite{science.aaa6486,PhysRevLett.115.126803}
In particular, Xenes made of group IV and V elements are referred to as silicene, germanene, stanene, plumbene, phosphorene, arsenene, antimonene, and bismuthene when X=Si, Ge, Sn, Pb, P, As, Sb, and Bi, respectively. In contrast to the planar configuration of graphene, Xenes prefer a structural buckling that lower the energy. The structural stability of these buckled structures was usually confirmed by phonon calculations that show no imaginary frequency. However, the band topology of the phonon spectrum in these materials is ignored in previous studies.

In this work, we reveal the nontrivial phononic SW topology in a large class of Xenes and their ligand-functionalized derivatives. Based on first-principles calculations, we identify the nontrivial phononic SWIs by directly calculating the second SW number $w_2$ of band gaps in the phononic spectrum and topological corner modes on nanodisk samples. By analyzing the vibrational modes at high-symmetry points, we find that the phononic SWIs are driven by double band inversion between the in-plane acoustic and out-of-plane optical modes with opposite parity eigenvalues in the presence of inversion symmetry. Due to the simple phononic structure and reliable synthesizability, our discovered phononic SWI materials provide an ideal platform to explore phononic SW topology.

\begin{figure*}
\includegraphics[width =0.9\textwidth]{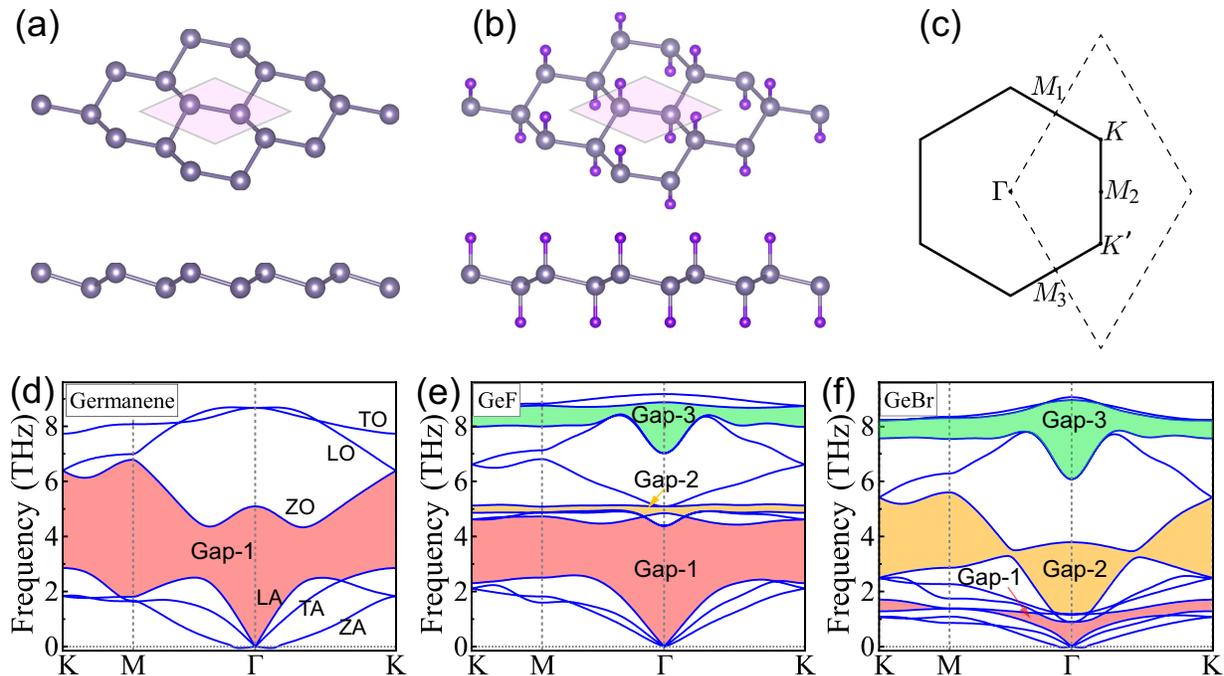}%
\caption{\label{fig1} (a,b) Crystal structure for 2D pristine Xenes (e.g., germanene) and ligand-functionalized Xenes (e.g., hydrogenated germanene GeH or halogenated germanene GeF and GeBr) from the perspective (side) view [upper (lower)]. Gray and purple spheres represent group-IV (Si, Ge, Sn) 
atoms and the ligands (-H, -F, -Cl, -Br), respectively. (c) Brillouin zone with high symmetry points labelled. (d-f) Phonon spectrum of pristine germanene, GeF and GeBr. The out of plane, transverse and longitudinal acoustic and optical modes are denoted by ZA, TA, LA, ZO, TO and LO, respectively. The phonon band gaps are labeled as Gap-$n$ ($n=1,2,3$) with increasing frequency.}
\end{figure*}

\section{Method}
{The self-consistent electronic calculations are performed within the framework of density functional theory using the Vienna Ab initio Simulation Package (VASP).\cite{VASP} We use the projector-augmented-wave potential,\cite{PAW} the Perdew-Burke-Ernzerhof exchange-correlation functional,\cite{PBE} the plane-wave basis with an energy cutoff of 500 eV, and a $25\times 25\times 1$ Monkhorst-Pack $k$-grid in all calculations. Periodic slab approach was employed to model 2D materials and a vacuum layer of 15~\AA~was adopted to eliminate the spurious interaction between neighboring slabs. The crystal structures were fully relaxed until the forces on each atom were smaller than 0.001 eV/\AA. We carry out lattice dynamics calculations using the density functional perturbation theory to generate real-space interatomic force constants within a $7 \times 7 \times 1$ supercell. The phonon dispersions were calculated from the force constants using PHONOPY code.\cite{TOGO20151} To reveal topological features of phononic SW insulators in these materials, we constructed a Wannier tight-binding Hamiltonian of phonons from real-space interatomic force constants.\cite{WU2018405}
}

\section{Results and discussion}
\paragraph*{Atomic structure and phonon spectrum.}
The atomic structure of 2D pristine Xenes is shown in Fig.~\ref{fig1}(a), which are comprised of group-IV atoms (Si, Ge, and Sn) arranged in a buckled honeycomb lattice. The structural buckling leads to deviations away from a $sp^2$ hybridization in planar structures and enhances the overlap between $\pi$-bonding $p_z$ and $\sigma$ orbitals, which stabilizes their hexagonal structures. For liganded Xenes, the buckled group-IV honeycomb lattice is decorated by chemical functional groups, i.e., ligands such as hydrogen or halogens, alternatively on both sides. The covalently bound ligands directly couple to the half-filled $p_z$ orbitals in 2D Xenes, thereby removing $\pi$ bonding and resulting in a stable $sp^3$ hybridization. Both pristine and liganded Xenes belong to the space group $P\bar{3}m1$ (No. 164, $D^3_{3d}$), which contains the inversion symmetry $\mathcal{P}$, an in-plane twofold rotation axis $\mathcal{C}_{2x}$, and an out-of-plane threefold rotation axis $\mathcal{C}_{3z}$. Combined with the time-reversal symmetry $\mathcal{T}$, all these structures possess the space-time inversion symmetry $I_{ST}=\mathcal{PT}$ which is required for 2D SWIs. Since each unit cell of the pristine (liganded) Xene contains 2 (4) atoms, there should be a total of 6 (12) bands in their phonon spectrum, as discussed later.

Taking the buckled honeycomb monolayer of Ge, which is referred to as germanene, as example, we calculate the phonon spectra of the pristine and halogenated germanene (e.g., GeF and GeBr), as shown in Fig.~\ref{fig1}(d)-(f). It is found that the phonon spectrum spans a range of about 10 THz for pristine and halogenated germanene, except for hydrogenated germanene which extends the range up to about 60 THz (see Fig. S3 in the Supplementary Material \footnote{\label{fn}See Supplemental Material at http://link.aps.org/supplemental/xxx, for more details about the Wilson loop method, phonon dispersions and Wilson loop spectra of pristine and ligand-functionalized Xenes.}). There are three acoustic (A) branches and three/nine optical (O) branches for pristine/liganded germanene. Both acoustic and optical branches can be classified as transverse (T), longitude (L), and out-of-plane (Z) modes according to their vibrational displacements and the direction of wave propagation. Notably, one observes several phononic band gaps in the spectrum. The gap between acoustic and optical bands is referred to as Gap-1, and other gaps among optical bands are labeled as Gap-$n$ ($n=2,3,\cdots$) successively with increasing frequency. Next, we will discuss the SW topology of each gap by considering phonon bands below that gap, in analogous to the electronic counterpart.

\begin{table}
\caption{\label{tab1} Irreps of phonon modes at $\Gamma$ and $M$ for germanene (Ge), GeH, GeF, and GeBr. Noted that $E$ represents a doubly degenerate irrep and the subscript $u/g$ denotes $-/+$ parity eigenvalues of the mode.}
\begin{tabular}{c|c|cccccccccccc}
  \hline
  \hline
    & $\Pi_i$ & 1 & 2& 3 & 4& 5& 6& 7& 8& 9& 10 &11&12 \\
  \hline
  Ge &$\Gamma$ &  $E_u$ & & $A_{2u}$ & $A_{1g}$ & $E_g$ &  & & & & & & \\
            & $M$ &   $B_g$ & $B_u$ & $A_g$ & $B_u$ & $A_g$ & $A_u$ & & & & & & \\
  \hline
  GeH & $\Gamma$ & $E_u$ & & $A_{2u}$ & $A_{1g}$ & $E_g$& & $E_u$ & & $E_g$ &  & $A_{1g}$ & $A_{2u}$ \\
      & $M$ & $A_g$ & $B_g$ & $B_u$ & $B_u$ & $A_g$ & $A_u$& $B_g$ & $B_u$ & $A_g$ & $A_u$  & $B_u$ & $A_g$ \\
  \hline
  GeF & $\Gamma$ & $E_u$ & & $A_{2u}$ & $E_g$& & $A_{1g}$& $E_u$ & & $E_g$ &  & $A_{1g}$ & $A_{2u}$ \\
      & $M$ & $B_g$ & $A_g$ & $B_u$ & $A_g$ & $A_u$ & $B_u$& $B_g$ & $B_u$ & $A_g$ & $A_u$  & $B_u$ & $A_g$ \\
  \hline
  GeBr & $\Gamma$ & $E_u$ &  & $A_{2u}$ & $E_u$ & & $E_g$ & & $A_{1g}$ & $E_g$ &  & $A_{2u}$ & $A_{1g}$ \\
      & $M$ & $B_g$ & $A_g$ & $B_u$ & $A_u$ & $B_g$ & $A_g$ & $B_u$& $B_u$ & $A_g$ & $A_u$ & $A_g$ & $B_u$ \\
  \hline
  \hline
\end{tabular}
\end{table}

\paragraph*{Phononic SW topology.}
{We first focus on Gap-1, which separates the acoustic and optical bands.
For phononic systems preserving the $\mathcal{PT}$ symmetry, we have $(\mathcal{PT})^2=1$, which means that all the phonon eigenmodes can be made real-valued and the abelian Berry curvature vanishes at every momentum. Instead, the second-order band topology of these phononic systems can be characterized by the second SW number $w_2$.
If the system has inversion symmetry $\mathcal{P}$ additionally, the parity of phonon modes are well defined at time-reversal invariant momenta (TRIM, labeled as $\Lambda_i$),\cite{parity} which are the $\Gamma$ and three $M$ points in the 2D Brillouin zone [see Fig.~\ref{fig1}(c)]. More importantly, the second SW number $w_2$, which characterizes the SW topology, can be determined by parity eigenvalues,\cite{PhysRevB.99.235125,Ahn_2019,lee2020two,PhysRevLett.121.106403,PhysRevLett.123.216803}
\begin{equation}
(-1)^{w_2}=\prod_{i=1}^{4}(-1)^{\lfloor N^-_{n}(\Lambda_i)/2\rfloor},\label{w2parity}
\end{equation}
where $N_n^-(\Lambda_i)$ represents the number of modes below Gap-$n$ with negative parity at TRIM $\Lambda_i$. $\lfloor\cdot\rfloor$ is the floor function.
Table~\ref{tab1} lists the irreducible representations (irreps) of phonon modes at high-symmetry points (i.e., $\Gamma$ and $M$). For both pristine germanene and liganded grermanene (e.g., GeH, GeF and GeBr) the irreps of the three acoustic modes below Gap-1 are composed of a 2D $E_u$ and a 1D $A_{2u}$ irreps at $\Gamma$, which become $B_g, A_g$ and $B_u$ irreps at $M$. Given the convention that the subscript $u/g$ of irreps represent negative/positive parity eigenvalues, it is straightforwardly found that $N_1^-(\Gamma)=3$ and $N_1^-(M)=1$. Therefore, the second SW number of Gap-1 is $w_2=1$, indicating they are phononic SWIs.
Physically, the mismatch $N_1^-(\Gamma)-N_1^-(M)=2$ signals a double band inversion, which provides an intuitive picture to understand the nontrivial topology.\cite{PhysRevLett.121.106403,PhysRevLett.123.186401,PhysRevLett.126.066401,Hsu13255,xiao2021first,huang2021generic}
Since the trivial atomic limit with local vibrations at isolated atoms should have the same parity representations at all TRIMs, the double band inversion implies an obstructed atomic limit which cannot adiabatically connect to the trivial one.}

\begin{table}
\caption{\label{tab2} A summary of 2D material candidates for phononic SWIs. Frequency range of nontrivial phononic band gaps ($w_2=1$) are listed. The symbol $\surd$ denotes the nontrivial phononic gap without a global gap.}
\begin{tabular}{c|cc}
  \hline
  \hline
   2D materials & Gap-1 (THz) &Gap-2 (THz)\\
  \hline
  Silicene (Si)  & $\surd$ & \\
  Germanene (Ge) & $2.85-4.33$ & \\
  Stanene (Sn) & $1.56-3.00$ &  \\
  Blue phosphorene (P) & $7.26-11.62$& \\
  Blue arsenene (As) & $3.58-6.50$& \\
  $\beta$-antimonene (Sb) & $2.08-4.48$ & \\
  $\beta$-bismuthene (Bi) & $1.28-3.15$& \\
  CH  & $23.96-28.16$  & \\
  SiH & $5.93-10.05$  & \\
  SiF & $2.28-3.01$   & \\
  SiCl & $\surd$  &  $6.55-6.91$ \\
  SiBr & $\surd$  &  $\surd$ \\
  GeH  & $3.15-5.86$  & \\
  GeF  & $2.59-4.38$  & \\
  GeCl & $1.94-3.23$  & $4.59-4.97$ \\
  GeBr & $\surd$ &  $3.29-3.47$ \\
  SnH  & $1.91-3.64$ &  \\
  SnCl & $0.90-1.01$ & $2.60-2.62$ \\
  SnBr & $0.63-0.64$ & $2.17-2.36$ \\
  \hline
  \hline
\end{tabular}
\end{table}

{We further examine $w_2$ of Gap-1 for other 2D pristine and liganded Xenes. As presented in Table~\ref{tab2}, all of these Xenes and derivatives have a nontrivial SW number $w_2=1$ for the three acoustic bands which are separated from optical bands by Gap-1. In addition, we surprisingly found that other 2D materials with similar buckled honeycomb structures, such as blue phosphorene,\cite{} blue arsenene,\cite{} $\beta$-antimonene, and $\beta$-bismuthene,\cite{} are also phononic SWIs (see Table~\ref{tab2} and the Supplementary Material \footnotemark[\value{footnote}]). This indicates that the nontrivial phononic SW topology of Gap-1, which is a generic feature of these materials, should be directly related to their buckled honeycomb lattices. As a crosscheck, we also calculate the second SW number $w_2$ for all these materials using the Wilson loop method,\cite{PhysRevLett.121.106403,Ahn_2019,PhysRevX.9.021013} which further confirms their nontrivial SW topology (see the Supplementary Material \footnotemark[\value{footnote}]).}

\begin{figure}
\includegraphics[width =1\columnwidth]{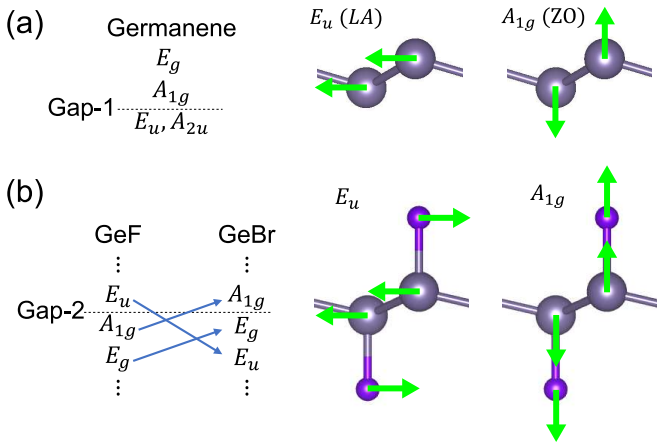}%
\caption{\label{fig2} Vibrational displacements of phonon modes at $\Gamma$ near (a) Gap-1 of pristine germanene and (b) Gap-2 of halogenated germanene (e.g., GeF and GeBr).}
\end{figure}

{Having demonstrated the nontrivial SW topology in these materials, we now analyze the underlying physical mechanism. By analyzing the vibrational modes of germanene [see Fig.~\ref{fig2}(a)], we found that the formation of Gap-1 stems from the hybridization between the out-of-plane $A_{1g}$ ZO mode and the in-plane $E_u$ LA mode due to the structural buckling. In comparison, the ZO mode of planar germanene has large imaginary frequencies at $\Gamma$ indicating instability. As the ZO mode corresponds to opposite displacements of Ge atoms alternatively along the z-axis, the imaginary part of the ZO mode would shift up and replace the portion of the LA mode that was connected to the LO mode upon structural buckling \cite{cahangirov2016introduction}. On the contrary, although the planar configuration is stable for graphene, the ZO mode of graphene overlaps with acoustic modes in frequency but does not hybrid with them to open the Gap-1. Therefore, structural buckling not only stabilizes the honeycomb structure of Xenes but also mixes the out-of-plane optical mode with in-plane acoustic modes, hence playing a crucial role in the realization of phononic SWIs.}

{Next, we consider other gaps in the phonon spectrum of liganded Xenes. By counting the number of phonon bands with the negative parity at $\Gamma$ and $M$ [i.e., $N_n^-(\Gamma)$ and $N_n^-(M)$] for each phonon band gap, we found that GeBr has a nontrivial SW number $w_2=1$ for Gap-2 but trivial $w_2=0$ for Gap-3, while GeF has trivial SW numbers for both gaps. The nontrivial topology of Gap-2 in GeBr is attributed to the band inversion between the double degenerate $E_u$ mode (LO and TO) and the $A_{1g}$ mode (ZO), which have opposite parity eigenvalues. As shown in Fig.~\ref{fig2}(b), the in-plane $E_u$ mode corresponds to relative shear displacements of Ge and halogen atoms, while the out-of-plane $A_{1g}$ mode corresponds to opposite displacements of atoms in AB sublattices. Because the atomic mass of Br (79.9 amu) which is comparable to Ge (72.6 amu) is much heavier than F (19.0 amu) and the Ge-Br bond (2.35\AA) is longer than the Ge-F bond (1.79\AA), the $E_u$ mode slows down in GeBr, which leads to frequency downshifting and the band inversion. It is worth noting that the number of bands below Gap-2 is 7 (6) for GeBr (GeF). Due to the negative parity of the 2D $E_u$ mode, $N_2^-(\Gamma)$ increases by 2 for GeBr, resulting in a nontrivial SW topology of Gap-2 according to Eq.~(\ref{w2parity}). The nontrivial Gap-2 of other liganded Xenes are induced by similar band inversion mechanisms between optical modes with opposite parties, hence are not elaborated here.}

\begin{figure*}
\includegraphics[width =1\textwidth]{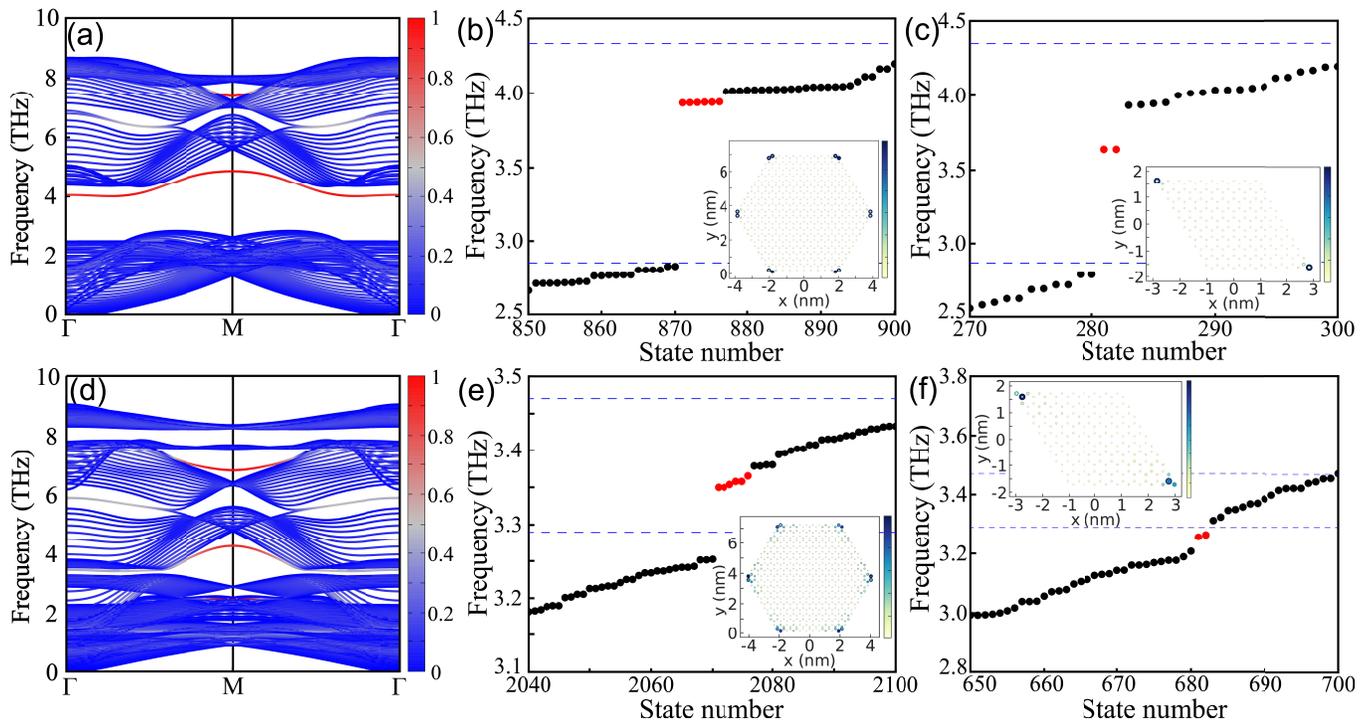}%
\caption{\label{fig3} Edge and corner states of (a-c) pristine germanene and (d-f) GeBr. (a, d) Phononic dispersion of a nanoribbon of (a) pristine germanene and (d) GeBr. The red (blue) represents edge (bulk) states. (b,e) Phononic spectrum of a hexagonal-shaped nanodisk near (b) the Gap-1 of pristine germanene and (e) the Gap-2 of GeBr, where the blue dashed line denotes the frequency range of the bulk phononic gap and the red dots denote corner vibrational modes. The inset shows the spatial distribution of corner modes, {where the size and color of circles denote the intensity of phonon mode.} (c,f) Same as in panels (b,e), but for diamond-shape samples.}
\end{figure*}

\paragraph*{Corner vibrational modes.}
Generally, the existence of gapped edge states and in-gap topological corner states serves as a characteristic feature of electronic SWIs due to the modified bulk-boundary correspondence. However, due to the lack of additional symmetry, such as chiral or particle-hole symmetry, the frequencies of topological corner vibrational modes are not strictly pinned in the bulk phononic gap. To further reveal the nontrivial SW topology, we study the edge and corner states by calculating phonon spectra of nanoribbons and nanodisks. As shown in Fig.~\ref{fig3}(a,d), there is one edge state inside the bulk Gap-1 of pristine germanene and Gap-2 of GeBr, which does not connect the upper and lower bands of the phononic gap. Moreover, there are six corner modes inside the bulk Gap-1 (Gap-2) for a hexagonal sample of pristine germanene (GeBr), as shown in Fig.~\ref{fig3}(b) and~\ref{fig3}(e). The spatial distribution of these phonon modes is localized at six corners (see inset of Fig.~\ref{fig3}b and \ref{fig3}e), verifying the character of in-gap topological corner states for 2D phononic SWIs. In addition, we also show the existence of corner vibrational modes in diamond-shape samples [Fig.~\ref{fig3}(c) and~\ref{fig3}(e)] and rectangular-shape samples (see Fig. S6\footnotemark[\value{footnote}]), indicating that the essential result is general for any sample geometry that preserves the $\mathcal{PT}$ symmetry.\cite{PhysRevB.104.085205,PhysRevB.105.085123}

Remarkably, it is found that corner vibrational modes are even robust against symmetry-breaking perturbations.\cite{PhysRevLett.119.246401} As an example, we break the inversion symmetry by randomly modifying the hopping parameters in the phononic tight-binding model of nanodisks, and find that the corner states maintain as long as the bulk and edge gaps are not closed, despite that their frequencies shift slightly (see Fig. S7 in the Supplementary Material \footnotemark[\value{footnote}]). Therefore, the corner states are indeed very robust and the lattice symmetry itself is not essential for their existences in our proposed phononic SWIs.

\paragraph*{Discussion.}
For a system with one atom per unitcell, the phonon spectrum which only contains 3 acoustic bands, cannot form any phononic gaps, not to mention the realization of phononic SWIs. Therefore, our proposed buckled honeycomb systems with only 2 atoms per unitcell (i.e., pristine Xenes made of group IV and V elements) should be the simplest phononic SWIs discovered ever. Moreover, the candidate materials with phononic gaps as large as 4.2 THz for hydrogenated graphene (referred to as graphane, CH in Table~\ref{tab2}) will facilitate experimentally detecting in-gap corner modes.
Since only a few phononic SWI materials (e.g., graphyne and graphdiyne\cite{acs.nanolett.1c04239,PhysRevB.105.085123}) with complex phononic spectra have been predicted so far, our proposal greatly extends the range of candidate materials for realizing phononic SWIs, which is expected to draw immediate experimental attention.

Notably, some of our proposed phononic SWI materials have been successfully synthesized in experiments. \cite{mannix2017synthesis, Grazianetti2020xenes, ANTONATOS2020100502,zhang2021recent} For example, 2D Xenes including silicene,\cite{PhysRevLett.108.155501,Liu_2014} germanene,\cite{D_vila_2014,nn4009406,jiang2014improving} and stanene \cite{zhu2015epitaxial} have been recently realized by molecular beam epitaxy growth or mechanical exfoliation. Moreover, the hydrogenation of graphene \cite{elias2009control, balog2010bandgap, REVgraphane} and germanene \cite{bianco2013stability} can be controlled accurately. In addition, buckled group-V honeycomb monolayers (e.g., blue phosphorene,\cite{smll.201804066} blue arsenene,\cite{Shah_2020} $\beta$-antimonene,\cite{ji2016two,wu2017epitaxial,zhu2019evidence} and bismuthene\cite{science.aai8142}) also have been grown by van der Waals epitaxy on substrates.
We, therefore, expect that it is experimentally feasible to detect the exotic phononic SWI states in these materials, e.g., by optical or neutron scattering.

\section{Conclusion}
In summary, we have shown phononic SWIs in a large, well-studied, and readily synthesizable class of 2D Xenes and their ligand-functionalized derivatives. We characterize the topological nature by the nontrivial second SW number $w_2=1$ {and topological corner vibrational modes}. Due to the structural buckling of their 2D honeycomb structure, the in-plane and out-of-plane phononic modes with opposite parities are coupled with each other, which realizes the double band inversion and results in the nontrivial phononic SW topology. {Our findings greatly extend the material candidates of phononic SWIs and provide useful guidance for the further investigation of phononic topological states.  The proposed phononic SWIs are expected to inspire future experimental studies, such as using inelastic neutron scattering or X-ray scattering spectroscopy to measure the topological phononic spectrum. Moreover, the corner vibrational modes can be excited by infrared light with resonant frequencies and detected by local probes such as scanning tunneling microscopy.}

\begin{acknowledgments}
This work was supported by the National Key R\&D Program of China (No. 2021YFA1401600), the National Natural Science Foundation of China (Grant No. 12074006), and the start-up fund from Peking University. The computational resources were supported by the high-performance computing platform of Peking University.
\end{acknowledgments}

\providecommand{\noopsort}[1]{}\providecommand{\singleletter}[1]{#1}%

\end{document}